\documentclass[%
 amsmath,amssymb,
 aps,
]{revtex4-2}

\usepackage{graphicx,subcaption,ragged2e}
\usepackage{dcolumn}
\usepackage{bm}
\usepackage{tikz}
\usepackage{graphicx}
\usetikzlibrary{positioning}

\usepackage{soul} 
\usepackage{float}  
\usepackage{caption}
\usepackage{subcaption}
\usepackage{hyperref}
\usepackage{multirow}
\usepackage{colortbl}
\usepackage{xcolor}

\begin{document}

\preprint{APS/123-QED}

\title{Turbulence Without the Viscous Tilting of Vorticity}
\author{Amr Emam\footnote{asabdelr@uci.edu},Mostafa Kamal\footnote{abdelmoh@uci.edu}, Perry L. Johnson\footnote{perry.johnson@uci.edu}}
\affiliation{%
 Department of Mechanical and Aerospace Engineering, University of California, Irvine
}%




\date{\today}

\begin{abstract}
Vortex stretching is a fundamental aspect of Navier-Stokes turbulence and is commonly understood in analogy to the stretching of infinitesimal material lines.  However, the parallel alignment of material lines and vorticity cannot be maintained due to the role of viscosity in the directional realignment of vorticity. In Navier-Stokes turbulence, the result is relatively modest quantitative differences in the alignment and stretching rates of vorticity compared to material lines. In this study, the qualitative effect of viscous tilting of vorticity is demonstrated directly by surgically removing it from direct numerical simulations of isotropic turbulence. The result is a drastic change to the fundamental structure of the flow, including a substantial deviation from the -5/3 inertial range scaling of the energy spectrum brought about by the infiltration and prevalence of viscous effects beyond the smallest scales. These observations demonstrate that the viscous tilting of vorticity is an essential characteristic of fluid turbulence. By extension, the same may be said of the difference in orientation and stretching rates for vorticity and infinitesimal material lines.
\end{abstract}

\maketitle

\section{Introduction}
\label{sec:introduction}

Following a fluid particle pathline through an incompressible flow, and neglecting the effect of viscosity, the vorticity magnitude is amplified by the same factor as the length of an infinitesimal material line parallel to the vorticity vector.
Taylor \cite{Taylor1938} appealed to this observation to explain the tendency of (three-dimensional) turbulent flows to amplify enstrophy by vortex stretching, rather than attenuate it via vortex compression.
Onsager \cite{Onsager1949} invoked Taylor's argument to explain the role of vortex stretching in the energy cascade from large to small scales in turbulence.
These considerations form a conceptual link between key universal features of turbulence across a wide variety of physical contexts: Lagrangian chaos, vortex dynamics, and high dissipation rates.
Over decades of investigations, a focus on vorticity and vortex dynamics in turbulence has remained prevalent, e.g., \cite{Tennekes1972, Chorin1988, Pullin1998, Eyink2008, Johnson2024, Buaria2026}.

The mathematical correspondence between vortex lines and infinitesimal material lines no longer holds for the viscous equations. Even when a material line is initialized parallel to the local instantaneous vorticity vector, the viscous evolution equations do not permit it to maintain parallel alignment with vorticity over time as the inviscid equations allow \cite{Guala2005}. After the initial alignment is broken, the active nature of vorticity (i.e., influencing the evolution of the strain-rate tensor and its eigenvectors) further distinguishes the evolution of vortex lines and material lines \cite{Guala2005}.

Several relevant observations about the effect of viscous tilting and the differences between the stretching of material lines and vorticity have previously been deduced from experiments and direct numerical simulation (DNS). The vorticity vector tends to preferentially align with the strain-rate eigenvector associated with its intermediate eigenvalue \cite{Ashurst1987, Tsinober1992}, which tends to be positive \cite{Lund1994}, and perpendicular to the most compressive eigenvalue. However, the vorticity aligns more closely with largest eigenvalue of (cumulative) material stretching than with the largest instantaneous strain rate eigenvalue \cite{Ni2014}. Comparatively, material lines show more proclivity to align with the largest strain-rate eigenvalue \cite{Johnson2016PRF}, and have modestly larger stretching rates \cite{Guala2005}. 
Earlier measurements of the inviscid contribution to the tilting of vorticity by Ref.\ \cite{Guala2005} indirectly suggested that the viscous contribution is strong.
However, later direct measurements of viscous tilting by Ref.\ \cite{Holzner2010} showed it to be typically significantly weaker than the inviscid tilting, although the two contributions were measured to be of comparable magnitude during events of large total vorticity tilting. 
Furthermore, the cumulative stretching of material lines is only modestly larger than the cumulative stretching of vorticity \cite{Johnson2016PRE}. 


In this paper, the effect of the viscous tilting of vorticity on turbulence is directly observed by surgically removing it from direct numerical simulations (DNS). While available evidence from previous studies tends to suggest that viscous tilting is responsible for relatively modest quantitative effects, the results of this study demonstrate that viscous tilting is central to fluid turbulence dynamics. Its removal leads to dramatic changes to the flow, including the elimination of -5/3 scaling range in the energy spectrum and the marginalization of inertial energy cascade mechanisms.

The remainder of the paper is structured as follows. Section \ref{sec:removal} explains the background, theory, and numerical implementation details for the removal of viscous tilting. The results are presented and discussed in Section \ref{sec:results}. In Section \ref{sec:conclusions}, implications of the study are discussed and conclusions are drawn.



\section{Removal of Viscous Tilting from Direct Numerical Simulations}
\label{sec:removal}

\subsection{Background \& Theory}
\label{sec:theory}

Juxtaposing the Lagrangian evolution of vorticity, $\omega_i = \epsilon_{ijk} \partial u_k / \partial x_j$, and an infinitesimal material line, $r_i$,
\begin{equation}
    \frac{d\omega_i}{dt} 
    = \frac{D\omega_i}{Dt} 
    = A_{ij} \omega_j + \nu \nabla^2 \omega_i,
    \hspace{0.1\linewidth}
    \frac{dr_i}{dt} = A_{ij} r_j,
    \label{eq:vorticity-vs-material-line}
\end{equation}
where $A_{ij} = \partial u_i / \partial x_j = S_{ij} - \frac{1}{2} \epsilon_{ijk} \omega_k$ is the velocity gradient tensor in relation to the vorticity and the strain-rate tensor, $S_{ij} = \tfrac{1}{2} \left( A_{ij} + A_{ji} \right)$.
Here, $D/Dt = \partial/\partial t + u_j \frac{\partial}{\partial x_j}$ is the material derivative.
Taylor's observation follows from Eq.\ \eqref{eq:vorticity-vs-material-line} when the flow is inviscid ($\nu = 0$) and the initial material line is parallel to the initial vorticity.

To elucidate the case of finite viscosity ($\nu \neq 0$), the vorticity and material line may be decomposed into magnitude and direction, $\omega_i = |\boldsymbol{\omega}| \tilde{\omega}_i$ and $r_i = |\mathbf{r}| \tilde{r}_i$, where the tilde denotes the unit vector of the associated quantity. Under this decomposition, the magnitudes evolve as,
\begin{equation}
    \frac{d\ln|\boldsymbol{\omega}|}{dt} = \tilde{\omega}_i \tilde{\omega}_j S_{ij} + \nu \tilde{\omega}_i \frac{\nabla^2 \omega_i}{|\boldsymbol{\omega}|},
    \hspace{0.06\linewidth}
    \frac{d\ln|\mathbf{r}|}{dt} = \tilde{r}_i \tilde{r}_j S_{ij}.
    \label{eq:magnitude-comparison}
\end{equation}
Both magnitudes grow according to the aligned rate of strain, while viscosity counteracts the growth of vorticity magnitude via enstrophy dissipation, proportional to the dot product between the vorticity and its Laplacian (observed to be preferentially, but not always, anti-parallel \cite{Holzner2010}).

Equation \eqref{eq:magnitude-comparison} highlights the sensitivity of stretching rates to the alignment with the strain-rate tensor. The unit vectors evolve as,
\begin{equation}
    \frac{d\tilde{\omega}_i}{dt} = 
    \left(\delta_{ik} - \tilde{\omega}_i \tilde{\omega}_k\right) A_{kj} \tilde{\omega}_j 
    + \nu \left(\delta_{ik} - \tilde{\omega}_i \tilde{\omega}_k\right) \frac{\nabla^2 \omega_k}{|\boldsymbol{\omega}|},
    \nonumber
    \hspace{0.06\linewidth}
    \frac{d\tilde{r}_i}{dt} =
    \left(\delta_{ik} - \tilde{r}_i \tilde{r}_k\right) A_{kj} \tilde{r}_j, ~
    \label{eq:unit-vector-comparison}
\end{equation}
where $\boldsymbol{\delta}$ denotes the Kronecker delta tensor. The inviscid reorientation of the vorticity and material lines by the velocity gradient (first RHS terms) supplies the mechanism for a bias toward alignment with positive strain rates \cite{Vieillefosse1982}, yielding positive magnitude growth on average in Eq.\ \eqref{eq:magnitude-comparison}. However, the viscous tilting of vorticity (second RHS term) has no counterpart in the material line evolution equation, providing a mechanism for differences in statistical alignment and stretching rates \cite{Holzner2010}.

Artificially removing the viscous tilting of vorticity leads to
\begin{equation}
    \frac{D\omega_i}{Dt}
    =
    A_{ij} \omega_j
    + \nu \tilde{\omega}_i \tilde{\omega}_j\nabla^2 \omega_j,
    \label{eq:vorticity-no-tilting}
\end{equation}
which can be solved in the same manner as Eq.\ \eqref{eq:vorticity-vs-material-line}, i.e., using Biot-Savart reconstruction of the velocity field at each time step.
That is, one may advance Eq.\ \eqref{eq:vorticity-vs-material-line} by finding the velocity using the Biot-Savart law (Poisson equation for the velocity),
\begin{equation}
   \nabla^2 u_i = - \epsilon_{ijk} \frac{\partial \omega_k}{\partial x_j},
\end{equation}
where the solution may be performed on Fourier modes,
\begin{equation}
    \hat{u}_i = \frac{\epsilon_{ijk} i \kappa_j \hat{\omega}_k}{|\boldsymbol{\kappa}|^2},
\end{equation}
where $\hat{\cdot}$ represents a Fourier transform and $\boldsymbol{\kappa}$ is the wavevector.

\subsection{Numerical Methods}
\label{sec:numerics}
The removal of viscous tilting from direct numerical simulations is facilitated by use of the vorticity form of the Navier-Stokes equations, Eq.\ \eqref{eq:vorticity-vs-material-line}. 
The vorticity transport equation is solved on a triply-periodic domain of size $(2\pi)^3$ via a pseudo-spectral method with low-wavenumber forcing (not shown in Eqs.\ \eqref{eq:vorticity-vs-material-line} and \eqref{eq:vorticity-no-tilting}) and phase-shift dealiasing \cite{Patterson1971}. 
Keeping the viscous tilting term, the solution produces DNS of stationary homogeneous isotropic turbulence consistent with previous velocity-based pseudo-spectral results using the same code \cite{Li2008, Johnson2020Energy, Johnson2021Role, Kamal2024}. A brief verification of our numerical implementation is shown in Appendix A.

Time advancement is accomplished with a second-order Adams-Bashforth scheme. 
The non-linear terms are dealiased using the $2\sqrt{2}/3$ rule with phase-shifting.
The body force is activated for only the first two wave-shells are forced at each time step with magnitude linearly proportional to the velocity. The coefficient of proportionality is determined at each time step to maintain constant kinetic energy spectrum for the first two wave-shells.
In this case the initial velocity field is chosen using a random number generator to create one instance from a Gaussian random field ensemble satisfying the divergence-free condition.
The initial velocity field is set to have a model energy spectrum, 
and the simulation is allowed to run for $6$ large-eddy turnover times to allow for the solution to become statistically representative of fully developed turbulence before statistics are recorded for $6$ more turnover times.
In both cases, simulations are performed over a range of Reynolds numbers $Re_{L}$ with grid resolutions from $64^3$ up to $1024^3$.

\begin{table*}[t]
\centering
\setlength{\tabcolsep}{6pt}
\begin{ruledtabular}
\begin{tabular}{c c c c c c c c c c}
Case & $N$ & $Re_\lambda$ & $Re_L$ & $\epsilon$ & $\nu$ &
$\eta^{*}$ & $\tau_\eta$ & $\Delta t$ & $k_{\max}\eta^{*}$ \\
\hline
Full DNS & $64^3$   & $57$  & $126$  & $0.096$ & $8.00\times10^{-3}$ & $4.20\times10^{-2}$ & $0.221$ & $4.0\times10^{-3}$ & $1.26$ \\
Full DNS & $128^3$  & $93$  & $300$  & $0.127$ & $3.20\times10^{-3}$ & $1.92\times10^{-2}$ & $0.115$ & $2.0\times10^{-3}$ & $1.15$ \\
Full DNS & $256^3$  & $155$ & $744$  & $0.107$ & $1.27\times10^{-3}$ & $9.72\times10^{-3}$ & $0.074$ & $1.0\times10^{-3}$ & $1.17$ \\
Full DNS & $512^3$  & $247$ & $1872$ & $0.111$ & $5.00\times10^{-4}$ & $4.78\times10^{-3}$ & $0.046$ & $5.0\times10^{-4}$ & $1.15$ \\
Full DNS & $1024^3$ & $396$ & $4654$ & $0.113$ & $2.00\times10^{-4}$ & $2.37\times10^{-3}$ & $0.028$ & $2.0\times10^{-4}$ & $1.14$ \\
\hline
No-Tilt DNS & $64^3$   & $-$  & $33$   & $0.364$ & $3.20\times10^{-2}$ & $2.80\times10^{-1}$ & $2.456$ & $4.0\times10^{-3}$ & $8.41$ \\
No-Tilt DNS & $128^3$  & $-$  & $116$  & $0.345$ & $9.00\times10^{-3}$ & $1.36\times10^{-1}$ & $2.045$ & $2.0\times10^{-3}$ & $8.14$ \\
No-Tilt DNS & $256^3$  & $-$  & $414$  & $0.391$ & $2.50\times10^{-3}$ & $6.61\times10^{-2}$ & $1.746$ & $1.0\times10^{-3}$ & $7.93$ \\
No-Tilt DNS & $512^3$  & $-$  & $1480$ & $0.392$ & $7.00\times10^{-4}$ & $3.33\times10^{-2}$ & $1.584$ & $5.0\times10^{-4}$ & $8.02$ \\
No-Tilt DNS & $1024^3$ & $-$ & $5189$ & $0.424$ & $2.00\times10^{-4}$ & $1.69\times10^{-2}$ & $1.423$ & $2.5\times10^{-4}$ & $8.13$ \\
\end{tabular}
\end{ruledtabular}
\caption{Numerical parameters for the full-DNS and no-viscous-tilting simulations. }
\label{tab:dns-simulation-parameters}
\end{table*}

\section{Results}\label{sec:results}

\subsection{Enegry Spectra}

\begin{figure}[t]
\begin{minipage}{0.5\textwidth}
  \begin{tikzpicture}
  \node (img)  {\includegraphics[scale=1]{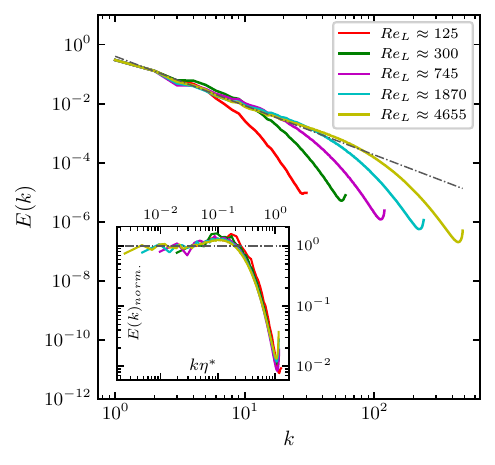}};
    \node[above=of img, node distance=0cm, xshift=-3.5cm, yshift=-1.5cm,font=\color{black}] {(a)};
  \end{tikzpicture}
\end{minipage}%
\begin{minipage}{0.5\textwidth}
  \begin{tikzpicture}
  \node (img)  {\includegraphics[scale=1]{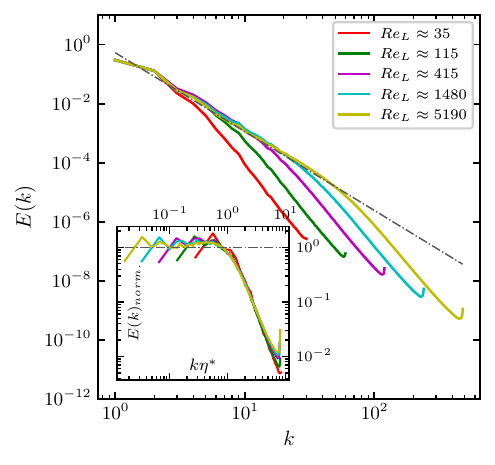}};
  \node[above=of img, node distance=0cm, xshift=-3.5cm, yshift=-1.5cm,font=\color{black}] {(b)};
  \end{tikzpicture}
\end{minipage}%
\caption{ Energy spectra comparison: 
  (a) DNS \cite{Kamal2024} and (b) DNS without viscous tilting. An infinite $Re$ spectrum
is shown for reference in both panels: (a) $E(k) \sim k^{-5/3}$ and (b) $E(k) \sim k^{-8/3}$. The inset in both panels shows the premultiplied spectrum.
}
  \label{fig:energy-spectra}
\end{figure}

Figure \ref{fig:energy-spectra} shows a side-by-side comparison of the energy spectra from the DNS with and without the viscous tilting of vorticity. In both cases, simulations are performed over a range of Reynolds numbers with grid resolutions from $64^3$ up to $1024^3$. The figure legends report the large-scale Reynolds number,$Re_L={u'}_{\mathrm{rms}} L/\nu$, rather than the Taylor-scale value, since the latter is tied to assumptions about the dissipation rate that are not as relevant when viscous tilting is removed. The range of Reynolds numbers is chosen to have the same effective range of length scales with and without viscous tilting. For the unaltered Navier-Stokes dynamics, Figure \ref{fig:energy-spectra}(a), the range of Taylor-scale Reynolds numbers is $Re_\lambda = 57$ up to $354$. As the Reynolds number is increased, the Kolmogorov length scale decreases and the inertial scaling range with $E(k) \sim k^{-5/3}$ emerges. The inset shows the premultiplied spectra, ${u'}_{\mathrm{rms}}^{-2} L^{2/3}  k^{5/3} E(k)$ against Kolmogorov-normalized wavenumber, highlighting the expected collapse of the high-wavenumber spectra across Reynolds numbers. In addition to the inertial scaling, the well-documented spectral bump associated with the bottleneck effect \cite{Johnson2021Role,donzis2010bottleneck, mininni2008nonlocal, frisch2008hyperviscosity, Agrawal2020, kurien2004cascade, Kamal2024, Arun_Kamal_Colonius_Johnson_2025} is also observed.

In Figure \ref{fig:energy-spectra}(b), the results without viscous tilting are shown. There is again a power-law scaling over an intermediate range of scales that emerges and expands with increasing Reynolds number. However, the observed power law is much steeper than $-5/3$ (approximately $-8/3$). This observed scaling is consistent with a H\"older exponent (more precisely, a 2nd-order Besov exponent \cite{Eyink2008}) of $5/6$. That is, $\langle \delta u_\ell^2 \rangle \sim \ell^{\zeta_2}$, where $\tfrac{1}{2}\zeta_2 = 5/6$.
This permits a modified definition of the viscous length scale from classical Kolmogorov theory ($\zeta_2 = 2h$), enforcing
$Re_\eta = \delta u_\eta \eta / \nu = 1$,
\begin{equation}
    \eta^{*} \equiv \eta_{spectra} = \left(\frac{\nu L^h}{u'_{\mathrm{rms}}}\right)^{1/(h+1)}.
    \label{eq:dissipation-scale}
\end{equation}
Note that for the standard Kolmogorov-based exponent, $h=1/3$, this definition differs from the commonly-used $\eta = \nu^{3/4} \epsilon^{-1/4}$ merely by a multiplicative factor, assuming the well-attested zeroth law of turbulence, $\epsilon = C_\epsilon (u_{\mathrm{rms}}^\prime)^3 L^{-1}$.

The inset of Figure \ref{fig:energy-spectra}(b) shows the premultiplied energy spectra, ${u'}_{\mathrm{rms}}^{-2} L^{5/3} k^{8/3} E(k)$, as a function of normalized wavenumber $k \eta^{*}$ according to Eq.\ \eqref{eq:dissipation-scale} with the empirical $h=5/6$. The resulting normalizations collapse the spectra across different wavenumbers, confirming that the small-scale cutoff is still related in some way to viscous dissipation. (Note that the inset of Fig. \ref{fig:energy-spectra}(a) also uses Eq.\ \eqref{eq:dissipation-scale}, but with $h=1/3$.) 
The deviation from classical Kolmogorov scaling suggests that the inviscid (inertial) nature of the intermediate range of scales no longer holds. That is, the removal of viscous tilting causes the viscous term to directly impact the dynamics in the intermediate range of scales $\ell \gg \eta$.


The impact of viscous tilting removal on larger scales, $\ell \gg \eta$, can be conceptually understood as follows. 
The velocity field evolves according to,
\begin{align}
    \frac{Du_i}{Dt}
    =-\frac{\partial (p/\rho)}{\partial x_i}&+T^*_i +f_i,
    \label{eq:NS}
\end{align}    
where $p$ is the pressure, $\rho$ is the density, $\mathbf{f}$ is the low-wavenumber forcing, and $\mathbf{T^*}$ is the viscous force. Navier-Stokes is recovered for $T_i = \nu \nabla^2 u_i$, where $\nu$ the kinematic viscosity. 
The curl of Eq.\ \eqref{eq:NS} recovers Eq.\ \eqref{eq:vorticity-vs-material-line} or \eqref{eq:vorticity-no-tilting} for the vorticity, where
\begin{equation}
    (\nabla\times T^*)_i
    = \left\lbrace
    \begin{array}{c c}
        \nu \nabla^2\omega_{i} & \text{Navier-Stokes} \\
        \nu\widetilde{\omega}_i \widetilde{\omega}_j\nabla^2\omega_j & \text{without viscous tilting}
    \end{array}
    \right..
    \label{eq:curl-of-viscous-force}
\end{equation}
Although the viscous force $\mathbf{T^*}$, for the case without viscous tilting, is not in general divergence-free, the Helmholtz-Hodge decomposition of $\mathbf{T^*}$ \
\begin{equation}
    \mathbf{T^*}= \nabla\Phi + \mathbf{T}
    \label{eq:Helmholtz-Hodge-decomposition}
\end{equation}
recovers the divergence-free property of the viscous force ($\nabla \cdot \mathbf{T} = 0$), because the curl-free part can be combined into a modified pressure $\nabla p^\prime=\nabla( p/\rho+\Phi)$ enforcing the divergence-free condition on the velocity field, $\nabla \cdot \mathbf{u} = 0$. Thus, $\nabla \times \mathbf{T} = \nabla \times \mathbf{T}^*$ for Eq.\ \eqref{eq:curl-of-viscous-force} and the incompressible momentum conservation equation in Eq.\ \eqref{eq:NS} provides equivalent velocity field evolution for $\mathbf{T}$ and $\mathbf{T}^*$.

The well-known vector calculus identity,
\begin{equation}
    \nabla \times \left(\nabla \times \mathbf{T} \right) 
    = \nabla \left(\nabla \cdot \mathbf{T}\right)
    - \nabla^2 \mathbf{T}
\end{equation}
combined with $\nabla \cdot \mathbf{T} = 0$ and the curl of Eq.\ \eqref{eq:curl-of-viscous-force} produces a Poisson equation for the viscous force in the absence of viscous tilting,
\begin{equation}
    \nabla^2 T_i
    =
    \nu \epsilon_{ijk} \frac{\partial}{\partial x_j}\left( \widetilde{\omega}_k \widetilde{\omega}_m \nabla^2 \omega_m \right).
\end{equation}
The Green's function solution in free space is,
\begin{equation}
    T_i(\mathbf{x}) = - \iiint \frac{1}{4 \pi |\mathbf{x}^\prime - \mathbf{x}|} \nu \epsilon_{ijk} \frac{\partial}{\partial x_j}\left( \widetilde{\omega}_k \widetilde{\omega}_m \nabla^2 \omega_m \right) d\mathbf{x}^\prime.
\end{equation}
Integration by parts then produces a Biot-Savart law for the viscous force,
\begin{equation}
    T_i(\mathbf{x}) = \iiint 
    \frac{\varepsilon_{ijk} \left(x_j^\prime - x_j \right) \left( \nu \left.\tilde{\omega}_k \tilde{\omega}_m \nabla^2 \omega_m \right)\right|_{\mathbf{x}^\prime}}{4 \pi |\mathbf{x}^\prime - \mathbf{x}|^3}  
    d\mathbf{x}^\prime,
    \label{eq:viscous-force-no-tilting}
\end{equation}

Equation \eqref{eq:viscous-force-no-tilting} underscores the nonlocality inherent to the act of removing the viscous tilting of vorticity.
Before the viscous tilting is removed (i.e., Navier-Stokes), the viscous force per unit volume is $T_i = \nu \nabla^2 u_i$, which is a localized operator that acts selectively on high wavenumbers proportional to $|\boldsymbol{\kappa}|^2$.
In Navier-Stokes turbulence, the viscous force is mostly confined to act only at the smallest scales, in the viscous range, i.e., where the spectrum deviates from the inertial range scaling in Fig. \ref{fig:energy-spectra}(a).
At larger scales, the viscous force becomes small compared to the inertial term in Navier-Stokes, allowing for the inertial range of scales with Kolmogorov $-5/3$ power-law scaling to emerge based on an inertia-driven exchange of energy across scales (energy cascade).

After the viscous tilting of vorticity is removed, the viscous force per unit volume in the momentum conservation equation, Eq.\ \eqref{eq:NS}, is given by the nonlocal operator in Eq.\ \eqref{eq:viscous-force-no-tilting} with power-law spatial decay. Thus, the removal of viscous tilting in the vorticity equation allows the viscous force to act more significantly at larger scales, rather than being mostly confined to act in a narrow range of smallest scales. The influence of $\mathbf{T}$ on larger scales provides a plausible mechanism for deviation from classical $-5/3$ scaling of the spectrum observed in Figure \ref{fig:energy-spectra}(b).
It is not yet evident from this consideration, however, why a new power-law scaling emerges in the intermediate scales. To understand this better, we next explore the scale-wise redistribution of kinetic energy.


\subsection{Energy Cascade}

The effect of viscosity on intermediate length scales and multiscale interactions can be quantified in terms of scale-wise kinetic energy
using a low-pass spatial filter (with a Gaussian kernel),
\begin{equation}
    \overline{u}_i^\ell = G_\ell \star u_i,
    \hspace{0.1\linewidth}
    G_\ell(\mathbf{r}) = \exp\left(-\frac{|\mathbf{r}|^2}{2\ell^2}\right),
\end{equation}
where $\star$ denotes a 3D spatial convolution.
Previous scale-wise kinetic energy analysis in Navier-Stokes turbulence showed little sensitivity to the filter shape \cite{Johnson2020Energy}.
The kinetic energy associated with length scales larger than $\ell$ is $E^\ell = \tfrac{1}{2} \overline{u}_i^\ell \overline{u}_i^\ell$ and its evolution equation may be derived by filtering Eq.\ \eqref{eq:NS}, followed by a dot product with $\overline{u}_i^\ell$,
\begin{equation}
\frac{\partial E^\ell}{\partial t}
+ \frac{\partial \phi_j^\ell}{\partial x_j}
=
\overline{u}_i^\ell \overline{f}_i^\ell
-\Pi^{\ell} 
-\mathcal{E}^\ell-\Pi^{\ell}_{vis},
\label{eq:resolvedKE}
\end{equation}
where $\phi_j^\ell = \frac{1}{2}\overline{u}_i^\ell \overline{u}_i^\ell \overline{u}_j^\ell + \frac{1}{\rho}\overline{p}^\ell \overline{u}_i^\ell\delta_{ij} + \overline{u}_i^\ell \sigma_{ij}^\ell$ is the spatial flux of large-scale kinetic energy and $\Pi^{\ell}=-\overline{S}_{ij}^\ell \sigma_{ij}^\ell$ is the inter-scale flux of kinetic energy by inertial mechanisms, i.e., the energy cascade rate due to vortex stretching and strain-rate self-amplification \cite{Johnson2020Energy, Johnson2021Role}. The residual stress tensor is $\sigma_{ij}^\ell = \overline{u_i u_i}^\ell - \overline{u}_i^\ell \overline{u}_i^\ell$.

The direct viscous dissipation of energy at scales larger than $\ell$ is $\mathcal{E}^\ell$, and $\Pi_{vis}^\ell$ represents the sink of large-scale ($>\ell$) kinetic energy due to viscous interactions with scales smaller than $\ell$. In Navier-Stokes turbulence, $\mathcal{E}^\ell = 2\nu \overline{S}_{ij}^\ell \overline{S}_{ij}^\ell$ and $\Pi_{vis} = 0$ (due to the linearity of the viscous term). When the viscous tilting of vorticity is removed, these expressions become,
\begin{equation}
    \mathcal{E}^\ell = \overline{u}_i^\ell T_i(\overline{\boldsymbol{\omega}}^\ell)
\end{equation}
\begin{equation}
    \Pi_{vis}^\ell = \overline{u}_i^\ell [\overline{T}_i^{\,\ell}-T_i(\overline{\boldsymbol{\omega}}^{\,\ell})] 
\end{equation}
where $T_i(\overline{\boldsymbol{\omega}}^{\,\ell})$ is the viscous stress computed by substituting the resolved vorticity field, $\overline{\boldsymbol{\omega}}^{\,\ell}$, in Eq.\eqref{eq:viscous-force-no-tilting}, and $\overline{T}_i^{\,\ell}$ results from applying the spatial filtering after obtaining $T_i$ from Eq.\eqref{eq:viscous-force-no-tilting} based on the unfiltered vorticity. Note that the multiscale viscous interactions occur due to the nonlinearity of the modified viscous force, Eq.\ \eqref{eq:viscous-force-no-tilting}.

\begin{figure}[htbp]
\begin{minipage}{0.5\textwidth}
  \begin{tikzpicture}
  \node (img)  {\includegraphics[scale=1]{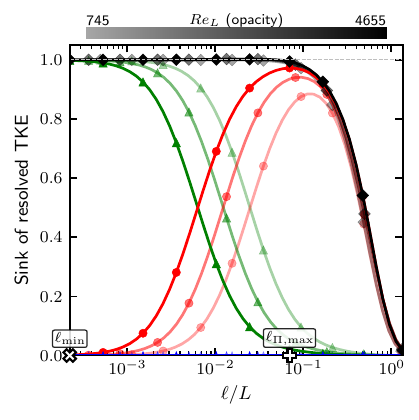}};
    \node[above=of img, node distance=0cm, xshift=-3.5cm, yshift=-1.5cm,font=\color{black}] {(a)};
  \end{tikzpicture}
\end{minipage}%
\begin{minipage}{0.5\textwidth}
  \begin{tikzpicture}
  \node (img)  {\includegraphics[scale=1]{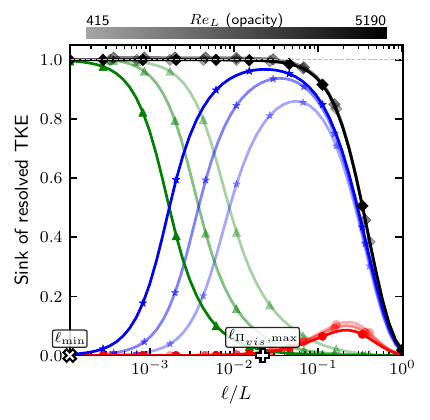}};
  \node[above=of img, node distance=0cm, xshift=-3.5cm, yshift=-1.5cm,font=\color{black}] {(b)};
  \end{tikzpicture}
\end{minipage}%

\begin{minipage}{1\textwidth}
 {\includegraphics[scale=1]{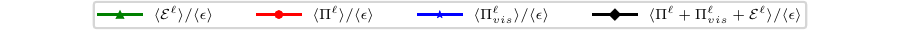}}
\end{minipage}%

\caption{For the three largest Reynolds numbers shown in Figure \ref{fig:energy-spectra}, the average inertial and viscous sinks of kinetic energy associated with length scales larger than $\ell$, based on Eq.\ \eqref{eq:resolvedKE}: (a) Navier-Stokes turbulence and (b) without viscous tilting.}
  \label{fig:energy-cascade}
\end{figure}

Figure \ref{fig:energy-cascade} shows a side-by-side comparison of the averages of the three sink terms on the RHS of Eq.\ \eqref{eq:resolvedKE} for the three largest Reynolds number simulations (Table \ref{tab:dns-simulation-parameters}), with and without viscous tilting. Also shown in black is the sum of all three sink terms, which balance the kinetic energy source term from the forcing, $\overline{u}_i^\ell \overline{f}_i^\ell$, due to stationarity. The results are normalized by the total rate at which kinetic energy is injected and dissipated, $\langle \epsilon \rangle = \langle u_i f_i \rangle$. Note that $\langle \overline{u}_i^\ell \overline{f}_i^\ell \rangle \approx \langle u_i f_i \rangle$ for $\ell \ll L$, where the black curves approach unity.
In Figure \ref{fig:energy-cascade}(a) for Navier-Stokes turbulence, the inertial energy cascade mechanism, $\Pi^\ell$, approaches equality with the dissipation rate for $\eta \ll \ell \ll L$ as the Reynolds number is increased. The viscous dissipation rate is confined to the scales near $\eta$, and thus shifts to the left as Reynolds number increases. This is the classical picture of hydrodynamic turbulence.

In Figure \ref{fig:energy-cascade}(b) for the case without viscous tilting of vorticity, it is the energy flux due to interscale viscous interaction, $\Pi_{vis}^\ell$, that approaches equality with the dissipation rate for $\eta \ll \ell \ll L$ as the Reynolds number increases. Meanwhile, the inertial energy cascade rate, $\Pi^\ell$, remains at $\lesssim 10\%$ of the dissipation rate and confined to a fixed range of large scales near $L$. The resolved viscous dissipation rate, $\mathcal{E}^\ell$, is dominant only for smallest scales, so it shifts to the left with increasing Reynolds number as in the Navier-Stokes case. 
It is evident from Figure \ref{fig:energy-cascade} that the multiscale viscous interactions activated by the removal of viscous tilting overwhelm the inertial energy cascade mechanism. As a result, the kinetic energy at large and intermediate scales is drained more rapidly than in Navier-Stokes. The result is the steepening of the energy spectrum and significantly smoothing the velocity field at intermediate scales.

\subsection{Flow Structure With and Without Viscous Tilting}

The inertial mechanism in Navier-Stokes turbulence relies on vortex stretching and strain-rate self-amplification, leaving an unmistakable signature in the statistics of (filtered) velocity gradients at small and intermediate scales \cite{Vieillefosse1982, Ashurst1987, Lund1994, Meneveau2011, Johnson2024}. 
The results presented so far show that the removal of viscous tilting significantly alters the mechanism the transfers energy from large to small scales, with the viscous-mediated transfer taking the dominant role rather than vortex stretching or strain-rate self-amplification.
This section explores in more detail the effect of viscous tilting removal on flow structure at small and intermediate scales.

\begin{figure}[t]
\begin{minipage}{0.5\textwidth}
  \begin{tikzpicture}
  \node (img)  {\includegraphics[scale=1]{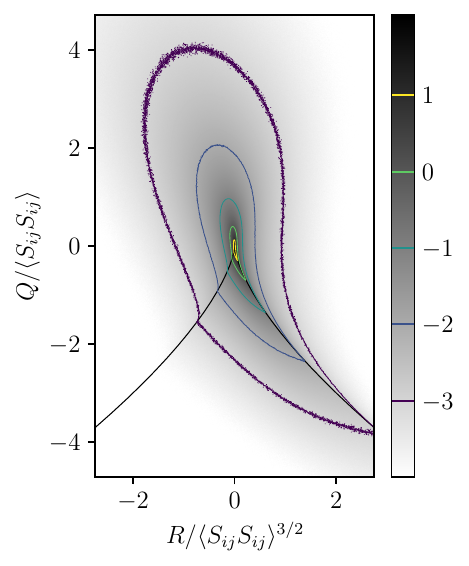}};
    \node[above=of img, node distance=0cm, xshift=-3.5cm, yshift=-1.5cm,font=\color{black}] {(a)};
  \end{tikzpicture}
\end{minipage}%
\begin{minipage}{0.5\textwidth}
  \begin{tikzpicture}
  \node (img)  {\includegraphics[scale=1]{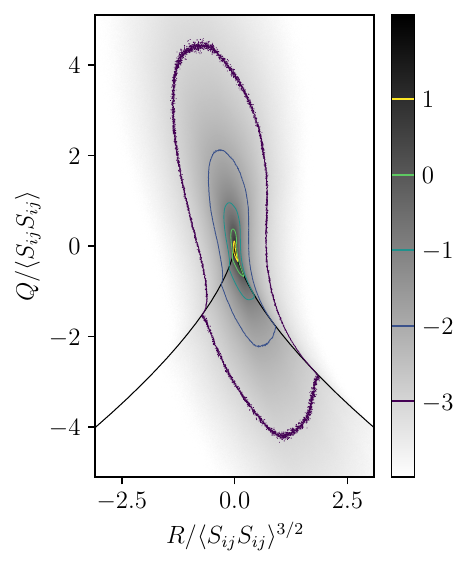}};
  \node[above=of img, node distance=0cm, xshift=-3.5cm, yshift=-1.5cm,font=\color{black}] {(b)};
  \end{tikzpicture}
\end{minipage}%
\caption{Joint PDFs of $Q$ and $R$, Eq.\ \eqref{eq:QR}, normalized by the mean square-magnitude of the strain-rate tensor, (a) Navier-Stokes turbulence, (b) without viscous tilting. In each respective case, the result from the largest Reynolds number simulation in Table \ref{tab:dns-simulation-parameters} is shown. The grayscale colorbar presents the base-10 logarithm of the PDF, and the colored contour lines represent the tick marks in the colorbar: $10^1$ (yellow), $10^0$ (green), $10^{-1}$ (blue-green), $10^{-2}$ (blue), $10^{-3}$ (purple).  }
  \label{fig:QR_PDF}
\end{figure}

The second and third invariants of the velocity gradient tensor,
\begin{equation}
    Q = -\tfrac{1}{2} A_{ij} A_{ji},
    \hspace{0.1\textwidth}
    R = -\tfrac{1}{3} A_{ij} A_{jk} A_{ki},
    \label{eq:QR}
\end{equation}
are useful quantities for characterizing the structure of small-scale turbulence \cite{Meneveau2011, Johnson2024}. 
Figure \ref{fig:QR_PDF} shows joint probability density function (PDF) for the (unfiltered) velocity gradients in Navier-Stokes turbulence (panel a) and in the absence of viscous tilting (panel b). The joint PDF for Navier-Stokes turbulence in panel (a) shows the well-known tear-drop shape, with large positive $Q$ events favoring $R < 0$ (vortex stretching dominated) states and large negative $Q$ events clustering along $R>0$ branch of the Viellefosse tail.

The corresponding results for the simulations with removal of viscous tilting shown in panel (b) have the same general features but with one noticeable difference. The shapes of the two innermost contours are largely similar. Progressing to more rare/extreme fluctuations in $Q$ and $R$, the contour lines show a modified shape for large negative $Q$ values. These extreme strain-rate dominated events are no longer tightly clustered along the $R>0$ the Viellefosse tail. However, there is still a lesser degree of asymmetry in the joint PDF. Overall, this suggests inertial mechanisms (e.g., as represented in the restricted Euler model \cite{Vieillefosse1982, Meneveau2011, Johnson2024}) are still influential but not as dominant in determining the flow structure at the smallest of scales.

\begin{figure}[t]
\begin{minipage}{0.5\textwidth}
  \begin{tikzpicture}
  \node (img)  {\includegraphics[scale=1]{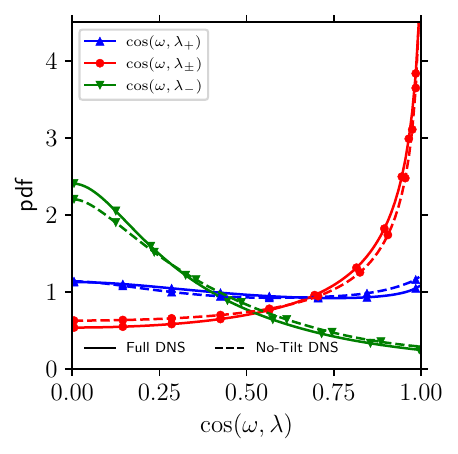}};
    \node[above=of img, node distance=0cm, xshift=-3.5cm, yshift=-1.5cm,font=\color{black}] {(a)};
  \end{tikzpicture}
\end{minipage}%
\begin{minipage}{0.5\textwidth}
  \begin{tikzpicture}
  \node (img)  {\includegraphics[scale=1]{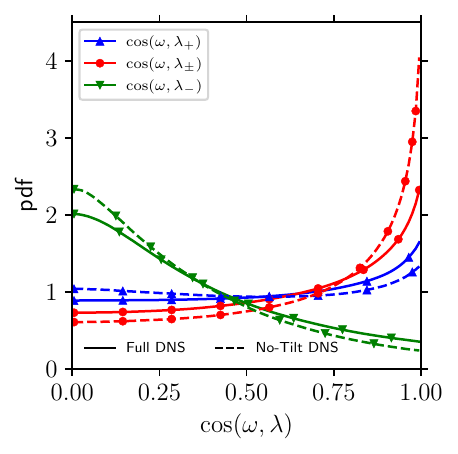}};
  \node[above=of img, node distance=0cm, xshift=-3.5cm, yshift=-1.5cm,font=\color{black}] {(b)};
  \end{tikzpicture}
\end{minipage}%
\caption{ PDFs of the alignment cosine between vorticity and the eigenvectors of the strain-rate tensor, with and without viscous tilting. The statistics are calculated from velocity gradients filtered (a) in the viscous range of scales, filtered at $\ell = \ell_{\min}$ indicated by $X$ symbols in Figure \ref{fig:energy-cascade}, and (b) in the intermediate range of scales, filtered at $\ell = \ell_{\Pi,\max}$ marked by $+$ symbols in Figure \ref{fig:energy-cascade}.}
  \label{fig:vorticity_alignment}
\end{figure}

Another striking but well-known feature of Navier-Stokes turbulence is the tendency of vorticity to preferentially align with the strain-rate eigenvector associated with the intermediate eigenvalue, $\lambda_{\pm}$, rather than that of the strongest positive ($\lambda_+$) or negative ($\lambda_-$) eigenvalue \cite{Ashurst1987, Meneveau2011, Johnson2024}. This fact is illustrated by the solid lines in Figure \ref{fig:vorticity_alignment}. Panel (a) shows results for the viscous range of scales, and panel (b) shows results in the intermediate range of scales. For Navier-Stokes turbulence, the preferential alignment of the vorticity with the intermediate strain-rate eigenvalue is stronger in the viscous range than in the inertial range of scales, cf.\ \cite{Fiscaletti2016}. This occurs together with a slight increase in the tendency to align with the strongest stretching eigenvector at intermediate scales.

The dashed lines in Figure \ref{fig:vorticity_alignment} show the results for the simulations without viscous tilting. At the smallest scales, Figure \ref{fig:vorticity_alignment}(a), there is remarkably little change in the alignments probabilities. In the intermediate range of scales, Figure \ref{fig:vorticity_alignment}(b), the alignment probabilities are largely unchanged from their small-scale counterparts. This results in modest quantitative differences between the Navier-Stokes and no-tilting cases at intermediate scales. Nevertheless, the same basic qualitative features of vorticity alignment in Navier-Stokes turbulence remain evident after the removal of viscous tilting.

Another well-known, salient feature of small-scale turbulence is the tendency of the strain-rate tensor to have two positive and one negative eigenvalue \cite{Lund1994, Meneveau2011, Johnson2024}. This can be quantified by inspecting the PDF of the $s^*$ parameter defined by Ref.\ \cite{Lund1994} as
\begin{equation}
    s^* = \Gamma_{SA} 
    = \frac{- \sqrt{6} S_{ij} S_{jk} S_{ki}}{\| \mathbf{S} \|^3} 
    = \frac{- 3 \sqrt{6} \lambda_{+} \lambda_{\pm} \lambda_{-}}{\left( \lambda_{+}^2 + \lambda_{\pm}^2 + \lambda_{-}^2 \right)^{3/2}},
    \label{eq:strain-rate-efficiency}
\end{equation}
which can be interpreted as the efficiency of strain-rate self-amplification \cite{Johnson2021Role, Kamal2024}, cf. Ref.\ \cite{Ballouz2018}. When calculated based on filtered strain rates in the intermediate range of scales, Eq.\ \eqref{eq:strain-rate-efficiency} represents the efficiency of the largest contributor to the inertial energy cascade, $\Pi^\ell$, shown in Figure \ref{fig:energy-cascade}. A similar metric for the efficiency of vorticity stretching can be defined as \cite{Johnson2021Role, Kamal2024},
\begin{equation}
    \omega^* = \Gamma_{VS}
    = \frac{\sqrt{6} ~\omega_i S_{ij} \omega_j}{2 |\boldsymbol{\omega}|^2 \| \mathbf{S} \|}
    = \frac{\sqrt{6} \left[ \lambda_{+} \cos^2(\omega,\lambda_{+}) + \lambda_{\pm} \cos^2(\omega, \lambda_{\pm}) + \lambda_{-} \cos^2(\omega, \lambda_{-}) \right]}{2\left( \lambda_{+}^2 + \lambda_{\pm}^2 + \lambda_{-}^2 \right)^{1/2}}.
    \label{eq:vorticity-efficiency}
\end{equation}
These two quantities are defined so as to be bounded between $-1 \leq \Gamma \leq 1$, consistent with their interpretation as efficiencies.

\begin{figure}[t]
\begin{minipage}{0.5\textwidth}
  \begin{tikzpicture}
  \node (img)  {\includegraphics[scale=1]{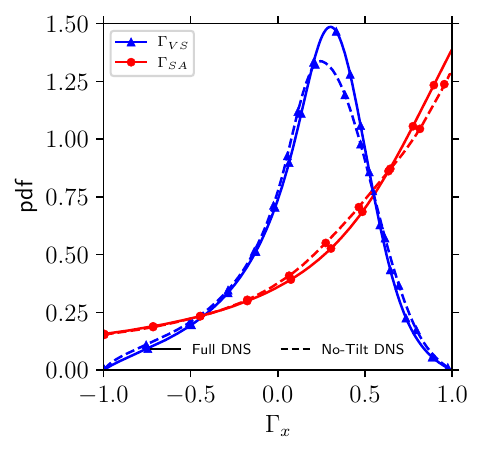}};
    \node[above=of img, node distance=0cm, xshift=-3.5cm, yshift=-1.5cm,font=\color{black}] {(a)};
  \end{tikzpicture}
\end{minipage}%
\begin{minipage}{0.5\textwidth}
  \begin{tikzpicture}
  \node (img)  {\includegraphics[scale=1]{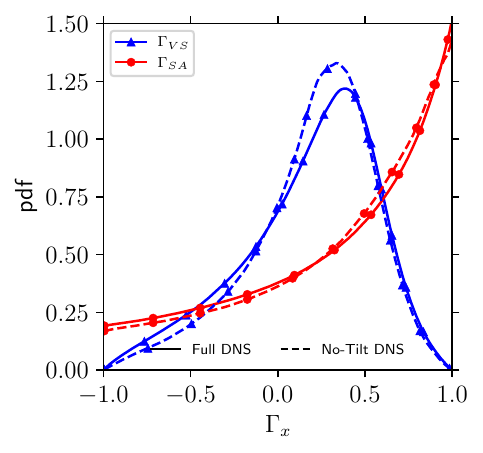}};
  \node[above=of img, node distance=0cm, xshift=-3.5cm, yshift=-1.5cm,font=\color{black}] {(b)};
  \end{tikzpicture}
\end{minipage}%
\caption{ PDFs of strain-rate self-amplification and vortex stretching efficiencies, as defined in Eqs.\ \eqref{eq:strain-rate-efficiency} and \eqref{eq:vorticity-efficiency}, respectively. The results are calculated from velocity gradients filtered (a) in the viscous range of scales, filtered at $\ell = \ell_{\min}$ indicated by X symbols in Figure \ref{fig:energy-cascade}, and (b) in the intermediate range of scales, filtered at $\ell = \ell_{\Pi,\max}$ marked by $+$ symbols in Figure \ref{fig:energy-cascade}.}
  \label{fig:efficiencies}
\end{figure}

The solid lines in Figure \ref{fig:efficiencies} show the distribution of these two efficiencies in Navier-Stokes turbulence, cf.\ Refs.\ \cite{Lund1994, Johnson2021Role, Kamal2024}. The strain-rate self-amplification efficience, $\Gamma_{SA}$, peaks at 100\% efficiency, which is characterized by two equal positive strain-rate eigenvalues and one negative eigenvalue with twice the magnitude. The most likely efficiency of vorticity stretching is less than 50\%, owing to the fact that vorticity does not achieve preferential alignment with the strongest strain-rate eigenvalue (Figure \ref{fig:vorticity_alignment}). Both efficiencies tend to be positive, indicating net enhancement of velocity gradients by each inertial mechanism. The distributions are similar in the viscous and inertial range, with minor quantitative differences.

The dashed lines in Figure \ref{fig:efficiencies} indicate the efficiency PDFs calculated from the simulations without viscous tilting. The efficiencies associated with inertial strengthening of gradients and the energy cascade are relatively unchanged by the removal of viscous tilting. This observation indicates that the relative change in the inertial cascade mechanism in Figure \ref{fig:energy-cascade} when the viscous tilting is removed is not due to changes in the alignment or structure of the turbulent fluctuations. Rather, it is due to the change in magnitude of the filtered gradients themselves. The weakening of the filtered gradient magnitudes is evident from the steeper spectra in Figure \ref{fig:energy-spectra}.

\begin{figure*}[h!]
\centering

\makebox[0.32\linewidth]{\hspace*{1.1cm}\textbf{$256^3$}}
\makebox[0.32\linewidth]{\hspace*{1.1cm}\textbf{$512^3$}}
\makebox[0.32\linewidth]{\hspace*{1.1cm}\textbf{$1024^3$}}

\includegraphics[width=0.32\linewidth]{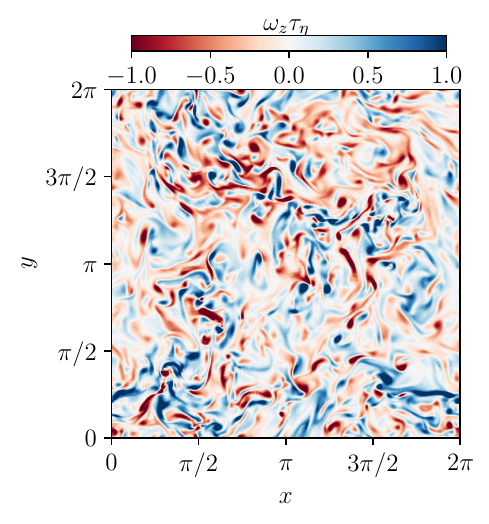}
\includegraphics[width=0.32\linewidth]{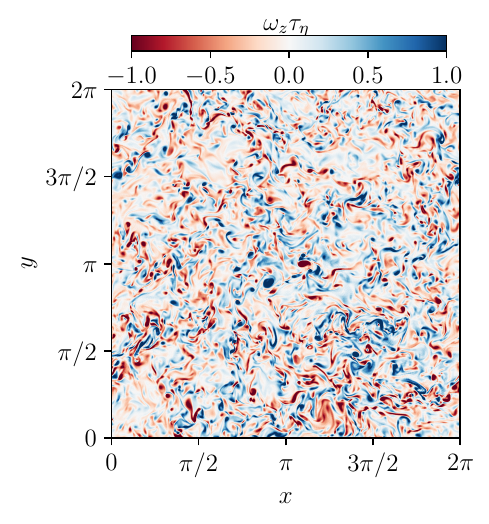}
\includegraphics[width=0.32\linewidth]{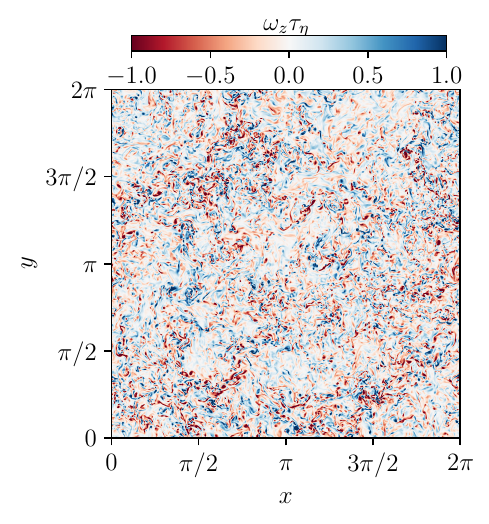}
\\[-0.5em]
\noindent\hspace*{1cm}\textbf{\footnotesize Full DNS}

\vspace{1em}

\includegraphics[width=0.32\linewidth]{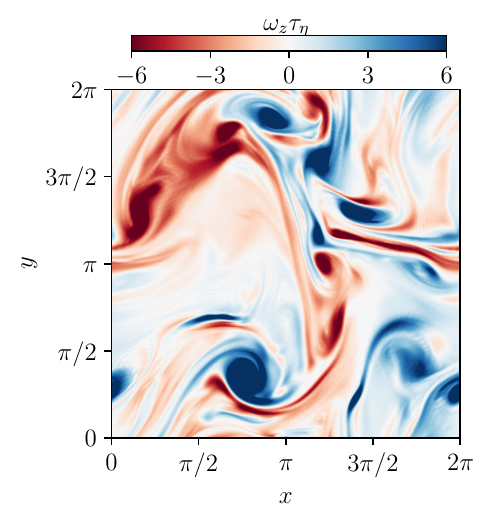}
\includegraphics[width=0.32\linewidth]{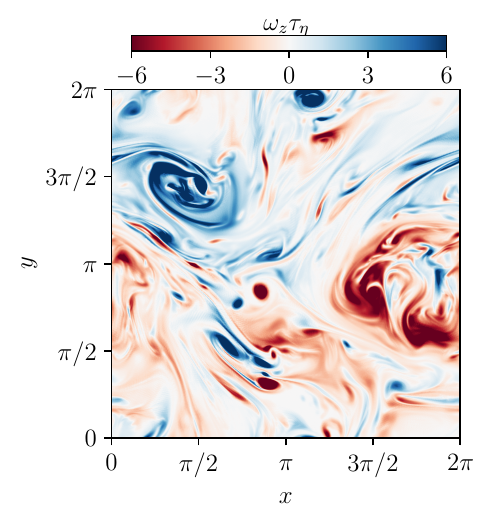}
\includegraphics[width=0.32\linewidth]{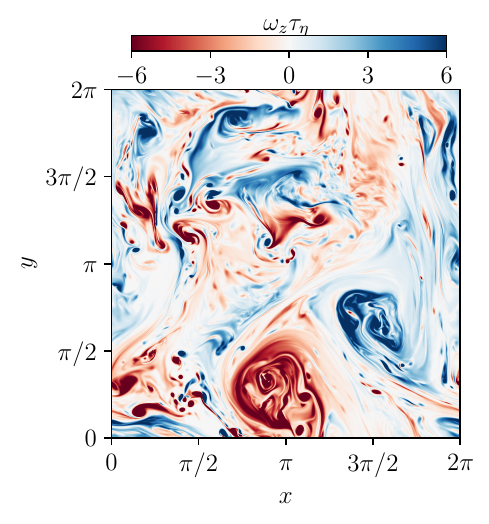}
\\[-0.5em]
\noindent\hspace*{1cm}\textbf{\footnotesize No-Tilt DNS}

\caption{Visualizations of the z-component of vorticity on x-y planes in simulations of Navier-Stokes turbulence (top), and simulations without the viscous tilting of vorticity (bottom). In both rows, the Reynolds number increases from left to right according to the simulation details recorded in Table \ref{tab:dns-simulation-parameters}.}
\label{fig:snapshots}
\end{figure*}

Finally, the flow structure with and without viscous tilting is visually inspected in Figure \ref{fig:snapshots}. In the top panels, the out-of-plane vorticity component is displayed from a two-dimensional plane in the three largest simulations of 3D Navier-Stokes listed in Table \ref{tab:dns-simulation-parameters}. As the Reynolds number increases, the vorticity structure is predominantly tied to the smallest (Kolmogorov) length scales, as is well documented.

Vorticity visualizations for a similar progression of increasing Reynolds number is shown from the simulations without viscous tilting in the lower panels of Figure \ref{fig:snapshots}. As the Reynolds number increases, smaller scale vortex tubes and sheets appear, reflecting the increasing range of scales. Unlike the Navier-Stokes dynamics, however, the simulations without viscous tilting maintain equally strong large-scale vortices. The simplest explanation is that the steeper spectra (approximately a $-8/3$ power-law) shown in Figure \ref{fig:energy-spectra} reflects an enstrophy spectrum with a $\approx -2/3$ power-law, close to a  $\kappa^{-1}$ power-law that would represent an equal distribution of intensity at all scales. By comparison, the typical Navier-Stokes enstrophy spectra power-law is $\kappa^{1/3}$, signifying a strong concentration of enstrophy magnitude at the smallest scales.

Synthesizing these observations about turbulence structure, the following picture emerges. The removal of the viscous tilting of vorticity from Navier-Stokes dynamics introduces a second mechanism for the transfer of energy from large to small scales, $\Pi_{vis}$, that is mediated by viscosity and not negligible in the intermediate range of scales.
This new viscous mechanism becomes dominant compared to the classical inertial mechanism for energy transfer (vorticity stretching and strain-rate self-amplification), leading to a steeper power-law energy spectrum and thus less energy at intermediate scales. Changes to the inertial cascade mechanism are minor in terms of flow structure, alignments, and efficiencies. Instead, the main reduction in the inertial mechanisms is due to the smaller amount of energy remaining in intermediate scales.

\section{Conclusions}\label{sec:conclusions}

In conclusion, the viscous tilting of vorticity is responsible for preventing it from maintaining parallel alignment with infintesimal material lines. When the viscous tilting is artificially removed from direct numerical simulations, dramatic changes occur. The inertial range with $E(k) \sim k^{-5/3}$ vanishes and is replaced by an intermediate range of scales with a much steeper power law (approximately $-8/3$) dominated by multiscale viscous interactions. This observation demonstrates that the viscous tilting of vorticity is essential to turbulent dynamics. Even though viscous tilting may seem to play a subtle quantitative role in Navier-Stokes dynamics, its removal leads to the catastrophic collapse of agreement with basic Kolmogorov theory. 
That is, artificially enforcing the alignment of vorticity and infinitesimal material lines causes a severe departure from basic turbulence scaling laws.
Therefore, the difference in orientation and stretching rates between material lines and vorticity must be regarded as an absolutely essential feature of turbulence.

These results underscore the need for a more nuanced understanding of the origin of (elevated levels of) vorticity via stretching than originally proposed by Taylor \cite{Taylor1938}. A helpful direction is provided by the generalization of Kelvin's theorem by Constantin \& Iyer \cite{Constantin2008}, which accounts for viscous effects via an ensemble of stochastic Lagrangian trajectories. The related adjoint-based formulation of Xiang, Eyink \& Zaki \cite{Xiang2025} provides a promising Eulerian approach for the back-in-time origin of vorticity in viscous flows. For example, the dynamics associated with generation and self-attenuation of extreme vorticity fluctuations at (viscous) sub-Kolmogorov scales has potentially important implications for open questions related to the regularity of Navier-Stokes \cite{Buaria2020, Buaria2024}.

Future work investigating the effect of diffusive tilting on other vector fields in a broader range of physical systems could lead to further insight, such as magnetic line stretching in magnetohydrodynamic (MHD) turbulence \cite{Capocci2025}. 
An analysis of the impact of removing diffusive tilting on other processes such as vortex reconnection could reveal other physical consequences.
Furthermore, eddy viscosity models for the residual stress tensor are common in large-eddy simulations (LES), although such models are known to have poor agreement with known strain-rate alignment statistics \cite{Ballouz2018}. A better understanding of the tilting of filtered vorticity caused by the residual stresses (as opposed to the molecular viscosity) appears important for developing more advanced models.

\section*{Acknowledgements}
This material is based upon work supported by the National Science Foundation under grant number 2152373, and the Office of Naval Research under grant number N000142512281.

\appendix

\section{Numerical Verification}

\label{app:verification}

\begin{figure}[h]
    \centering
    \includegraphics[width=0.495\linewidth]{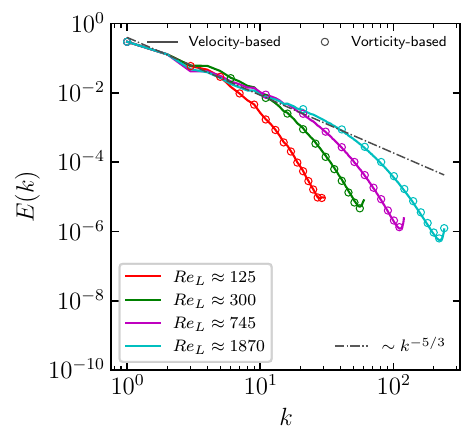}
    \caption{Verification of the vorticity-based DNS numerical method through comparison of energy spectra obtained from velocity- and vorticity-based simulations at different Reynolds numbers. 
    }
    \label{fig:dns-verification} 
    \label{fig:NT-resolution}
\end{figure}

This appendix presents a brief verification of the numerical methods used in this paper. Removing the viscous tilting of vorticity necessitated solving the Navier-Stokes equations in vorticity form, Eq.\ \eqref{eq:vorticity-vs-material-line}, instead of the typical velocity form. Figure \ref{fig:dns-verification}(a) compares the energy spectra for the vorticity-based Navier-Stokes implementation compared to that of a standard velocity-based pseudo-spectral implementation. The results are indistinguishable, verifying that the numerical methods implemented for solving Navier-Stokes in terms of the vorticity equation have the same accuracy as the original velocity-based method.


\bibliography{references}


\end{document}